\pdfoutput=1

\documentclass[8.5pt,twoside,twocolumn]{article}
\usepackage[usenames,dvipsnames]{xcolor}
\usepackage[super,sort&compress,comma]{natbib} 
\usepackage[version=3]{mhchem}
\usepackage{colortbl} 
\usepackage{sectsty}
\usepackage{balance} 
\usepackage[para,online,flushleft]{threeparttable}
\usepackage{graphicx} 
\usepackage{lastpage}
\usepackage[format=plain,justification=raggedright,singlelinecheck=false,font=small,labelfont=bf,labelsep=space]{caption} 
\usepackage{fancyhdr}
\definecolor{lightgray}{gray}{0.9}

\newcommand{\beginsupplement}{
        \setcounter{equation}{0}
        \renewcommand{\theequation}{S\arabic{equation}}%
        \setcounter{table}{0}
        \renewcommand{\thetable}{S\arabic{table}}%
        \setcounter{figure}{0}
        \renewcommand{\thefigure}{S\arabic{figure}}%
     }

\makeatletter
\newenvironment{tablehere}
{\def\@captype{table}}
{}
\newenvironment{figurehere}
{\def\@captype{figure}}
{}
\makeatother

\begin{document}

\thispagestyle{plain}
\fancypagestyle{plain}{
\renewcommand{\headrulewidth}{1pt}}
\renewcommand{\thefootnote}{\fnsymbol{footnote}}
\renewcommand\footnoterule{\vspace*{1pt}
\hrule width 3.4in height 0.4pt \vspace*{5pt}} 
\setcounter{secnumdepth}{5}

\makeatletter 
\def\subsubsection{\@startsection{subsubsection}{3}{10pt}{-1.25ex plus -1ex minus -.1ex}{0ex plus 0ex}{\normalsize\bf}} 
\def\paragraph{\@startsection{paragraph}{4}{10pt}{-1.25ex plus -1ex minus -.1ex}{0ex plus 0ex}{\normalsize\textit}} 
\renewcommand\@biblabel[1]{#1}            
\renewcommand\@makefntext[1]
{\noindent\makebox[0pt][r]{\@thefnmark\,}#1}
\makeatother 
\renewcommand{\figurename}{\small{Fig.}~}
\sectionfont{\large}
\subsectionfont{\normalsize} 

\fancyfoot{}
\fancyfoot[RO]{\footnotesize{\sffamily{1--\pageref{LastPage} ~\textbar  \hspace{2pt}\thepage}}}
\fancyfoot[LE]{\footnotesize{\sffamily{\thepage~\textbar\hspace{3.45cm} 1--\pageref{LastPage}}}}
\fancyfoot[RO]{\footnotesize{\sffamily{1--\pageref{LastPage} ~\textbar  \hspace{2pt}\thepage}}}
\fancyfoot[LE]{\footnotesize{\sffamily{\thepage~\textbar\hspace{2pt} 1--\pageref{LastPage}}}}
\fancyhead{}
\renewcommand{\headrulewidth}{1pt} 
\renewcommand{\footrulewidth}{1pt}
\setlength{\arrayrulewidth}{1pt}
\setlength{\columnsep}{6.5mm}
\setlength\bibsep{1pt}

\twocolumn[
  \begin{@twocolumnfalse}
\noindent\LARGE{\textbf{Solar energy conversion properties and defect physics of ZnSiP$_2^\dag$}}
\vspace{0.6cm}

\noindent\large{\textbf{Aaron D. Martinez,\textit{$^{a}$} Emily L. Warren,\textit{$^{b}$} Prashun Gorai,\textit{$^{a,b}$} Kasper A. Borup,\textit{$^{b,c}$}\\ Darius Kuciauskas,\textit{$^{b}$}  Patricia C. Dippo,\textit{$^{b}$} Brenden R. Ortiz,\textit{$^{a}$} Robin T. Macaluso,\textit{$^{d}$}\\ Sau D. Nguyen,\textit{$^{e}$} Ann L. Greenaway,\textit{$^{f}$} Shannon W. Boettcher,\textit{$^{f}$} Andrew G. Norman,\textit{$^{b}$}\\Vladan Stevanovi\'{c},\textit{$^{a,b}$} Eric S. Toberer,\textit{$^{a,b}$} and Adele C. Tamboli$^{\ast}$\textit{$^{a,b}$}}}\vspace{0.5cm}

\noindent\textit{\small{\textbf{Received Xth XXXXXXXXXX 20XX, Accepted Xth XXXXXXXXX 20XX\newline
First published on the web Xth XXXXXXXXXX 200X}}}

\noindent \textbf{\small{DOI: 10.1039/b000000x}}
\vspace{0.6cm}

\noindent
\normalsize{Implementation of an optically active material on silicon has been a persistent technological challenge. For tandem photovoltaics using a Si bottom cell, as well as for other optoelectronic applications, there has been a longstanding need for optically active, wide band gap materials that can be integrated with Si. ZnSiP$_2$ is a stable, wide band gap (2.1 eV) material that is lattice matched with silicon and comprised of inexpensive elements. As we show in this paper, it is also a defect-tolerant material. Here, we report the first ZnSiP$_2$ photovoltaic device. We show that ZnSiP$_2$ has excellent photoresponse and high open circuit voltage of 1.3 V, as measured in a photoelectrochemical configuration. The high voltage and low band gap-voltage offset are on par with much more mature wide band gap III-V materials. Photoluminescence data combined with theoretical defect calculations illuminate the defect physics underlying this high voltage, showing that the intrinsic defects in ZnSiP$_2$ are shallow and the minority carrier lifetime is 7 ns. These favorable results encourage the development of ZnSiP$_2$ and related materials as photovoltaic absorber materials.}

\vspace{0.5cm}

\colorbox{lightgray}{
\begin{minipage}{\textwidth-0.15in}
\section*{Broader Context}

Of all the renewable energy technologies, solar photovoltaic electricity has one of the highest resource potentials; there is enough energy in the sunlight incident on the surface of the earth to meet the world's energy demands many times over ($\sim$10,000:1). However, significant market penetration requires photovoltaics to be cost competitive with fossil fuels, even when unsubsidized. Currently, balance of system costs, rather than module costs, represent the majority of the total installed cost. Thus, increasing module efficiency is attractive as high efficiency cells can reduce installation size and therefore cost. Tandem photovoltaic architectures can provide a transformative boost in module efficiency over the single junction alternative due to reduced thermalization losses. Silicon photovoltaics is a well established ($>$90\% market share), high efficiency, low cost technology that provides a crystalline template to grow top cells upon. However, the top cell material must satisfy strict requirements, including high efficiency and long reliability, or its presence will simply reduce the performance of the silicon bottom cell. The primary top cell materials considered to date include III-V materials, but the cost of these materials and their sensitivity to defects have proven challenging. In this work, ZnSiP$_2$ emerges as a wide band gap absorber that has the potential to meet the requirements needed for a top cell in tandem silicon-based photovoltaics.

\end{minipage}
}
\vspace{0.5cm}
 \end{@twocolumnfalse}
  ]

\footnotetext{\dag~Electronic Supplementary Information (ESI) available:}


\footnotetext{\textit{$^{\ast}$~E-mail: Adele.Tamboli@nrel.gov}}
\footnotetext{\textit{$^{a}$~Department of Physics, Colorado School of Mines, Golden, Colorado 80401, USA}}
\footnotetext{\textit{$^{b}$~National Renewable Energy Laboratory, Golden, Colorado 80401, USA}}
\footnotetext{\textit{$^{c}$~Department of Chemistry, Aarhus University, Langelandsgade 140, DK-8000 Aarhus C, Denmark}}
\footnotetext{\textit{$^{d}$~Department of Chemistry and Biochemistry, University of Texas at Arlington, Arlington, Texas 76019, USA}}
\footnotetext{\textit{$^{e}$~Department of Chemistry, University of Northern Colorado, Greeley, CO 80639}}
\footnotetext{\textit{$^{f}$~Department of Chemistry and Biochemistry, Materials Science Institute, University of Oregon, Eugene, Oregon 97403, USA}}



\newpage

\begin{figurehere}
\centering
  \includegraphics[width=\columnwidth]{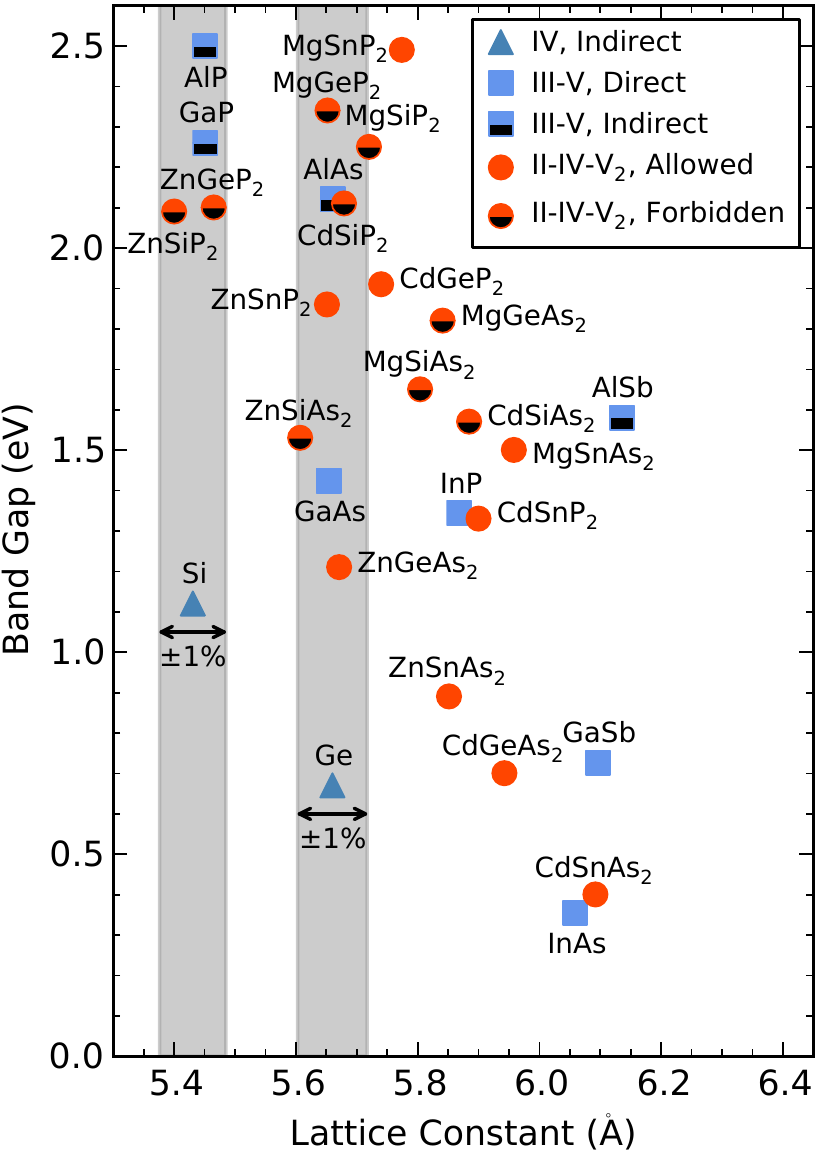}
  \caption{Theoretically determined band gaps versus lattice constants for some of the more earth abundant II-IV-V$_2$ chalcopyrites.  Also shown are III-V materials along with Si and Ge from group IV.  The gray vertical bars overlaying Si and Ge highlight materials with similar lattice constant, within $\pm$ 1\%.}
  \label{II-IV-V2}
\end{figurehere}

\section{Introduction}

Optoelectronic materials are typically semiconductors that are covalently bonded with tetrahedral coordination (i.e. group IV, II-VI and  III-V materials).  While the ternary II-VI analogs such as I-III-VI$_2$ (e.g.~CuInSe$_2$) have been extensively investigated,\cite{green2007} it is surprising that III-V analogs such as II-IV-V$_2$ (e.g.~ZnSiP$_2$) have received less attention for optoelectronic devices.\cite{kuhnel1975, shirakawa1979, shin1987, shin1989, wen1997, peshek2013, xing1991, choi2011} Figure \ref{II-IV-V2} shows that II-IV-V$_2$ chalcopyrite compounds span a broad range of band gaps, from 0.4~eV (CdSnAs$_2$) to 2.5~eV (MgSnP$_2$), and have lattice constants that are compatible with many unary and binary semiconductors. Many of the  II-IV-V$_2$ compounds are formed from comparatively inexpensive and non-toxic elements (e.g. Zn, Mg, Si, Sn, N, P), rendering them particularly attractive for applications requiring large scale deployment such as photovoltaics (PV). \cite{goodman1957, springthorpe1968, prochukhan1977, fioretti2015} Wide band gap II-IV-V$_2$ materials have seen virtually no investigation, despite significant opportunities for tandem PV, light emitting diode, photonic circuit, and laser applications.\cite{morkoc1994, green2001, soref2006} Within tandem PV, it has been challenging to find wide band gap materials suitable for tandems, and particularly those compatible with Si. \cite{Jain2014, Bailie2015} Thus, the II-IV-V$_2$  compounds ZnSiP$_2$ and ZnGeP$_2$ are of great interest as epitaxial top cell materials on a silicon bottom cell.   Implementing these materials as inexpensive, earth-abundant top cells on Si leverages the dominance of Si PV ($>$90\% market share).\cite{pvreport2014}  The work presented herein is focused on PV relevant characterization of ZnSiP$_2$.

The fundamental properties of ZnSiP$_2$ have been studied since the late 1950's using crystals that have been grown in a flux (typically Zn or Sn) or by halogen assisted vapor transport.\cite{springthorpe1968, springthorpe1970, ziegler1973, siegel1974, shay1975, gorban1976, siegel1976, prochukhan1977, averkieva1983, yao1986, buehler1971, shay1973, kuhnel1975, humphreys1975, humphreys1976, ziegler1978,  kuhnel1978, pamplin1979, folberth1958, clark1973, gentile1974} Studies of these crystals reveal that ZnSiP$_2$ has a very small lattice mismatch with Si of 0.5\% (Fig.~\ref{II-IV-V2}), has a band gap of $\sim$2.1~eV, forms with minimal atomic disorder, and is structurally stable at temperatures up to 800~$^{\circ}$C.\cite{martinez2015, folberth1958, gentile1974, pamplin1979,springthorpe1968, abrahams1970, bertoti1970, clark1973, prochukhan1977} Doping of ZnSiP$_2$ has yielded n-type (Se, Te, In, or Ga) and p-type (Cu) crystals. \cite{springthorpe1968, springthorpe1970, shay1973, ziegler1973, ziegler1978}

Some characterization has been done which is specifically relevant to the applications of tandem PV cells with silicon. Several authors have proposed and discussed the prospects of ZnSiP$_2$ heterojunctions with Si.\cite{prochukhan1977, averkieva1983, bachmann1993, schilfgaarde2010} Growth of Si/ZnSiP$_2$ interfaces has been demonstrated through heteroepitaxial crystallization of Si on ZnSiP$_2$ substrates,\cite{bertoti1970} growth of polycrystalline ZnSiP$_2$ on Si\cite{curtis1970}, and epitaxial ZnSiP$_2$ on Si by vapor-liquid-solid growth.\cite{popov1972}  While photoconductivity has been demonstrated, no PV devices have been realized to date.\cite{kuhnel1975, shirakawa1979, barreto1987}  A ZnSiP$_2$/Si device may be expected to have good light transmission through the top cell and into the bottom cell, because ZnSiP$_2$ has little parasitic below-band-gap absorption and good index of refraction matching with Si (reflection at the Si/ZnSiP$_2$ interface would be less than 1\%).\cite{martinez2015}  Considering that it is a stable, bipolar dopable, wide band gap semiconductor that is lattice and index of refraction matched with Si, ZnSiP$_2$ is a promising material for implementation as a top cell on Si PV.

In this work, we have employed a variety of theoretical and experimental techniques to address some of the challenges and unknowns regarding ZnSiP$_2$ as a top cell material on Si PV.  
Through a combination of photoluminescence measurements and first principles calculations, we show that the intrinsic material contains both donor and acceptor defects, and the associated energy levels are shallow ($\sim$0.1~eV from the associated band edge), resulting in defect-tolerant behavior. We identify the origin of these defects as antisite defects.
We show photoelectrochemical (PEC) measurements of ZnSiP$_2$ which demonstrate excellent photoresponse and high open circuit voltage (V$_\text{oc}$) of 1.3~V. Despite this being the first report of a ZnSiP$_2$ photovoltaic device, the observed voltage is higher than any other material lattice matched to Si, to our knowledge.\cite{Geisz2005, Song2015, Winkler2014}
These findings establish  ZnSiP$_2$ as a potential monolithic top cell material on Si.

\begin{figure}
\centering
  \includegraphics[width=\columnwidth]{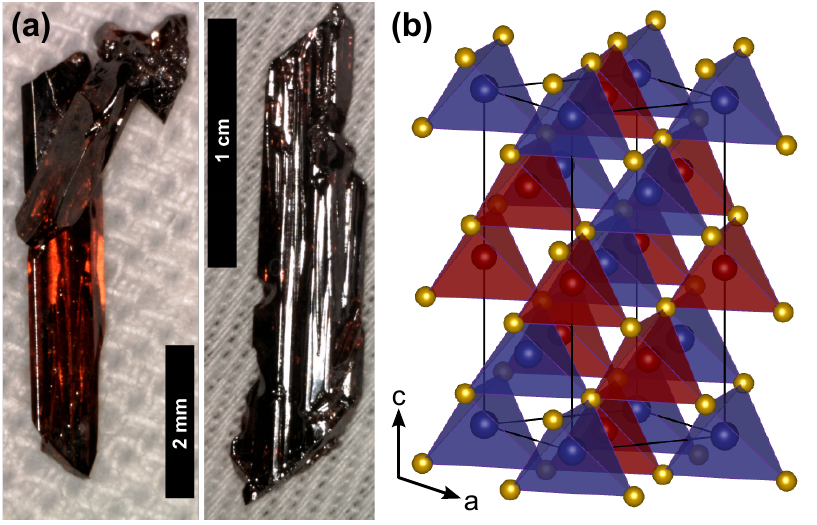}
  \caption{(a) Photograph of ZnSiP$_2$ single crystals grown in Zn flux. (b) Crystal structure of II-IV-V$_2$ chalcopyrites (I$_{\bar{4}2d}$) emphasizing two interlaced networks of corner sharing tetrahedra. In one network, the tetrahedra (red) have a Group IV atom at the center and Group V atoms on the corners. The second network differs from the first, in that the tetrahedra (blue) are slightly distorted and they have a Group II atom at the center.}
  \label{crystal}
\end{figure}

\section{Results and Discussion}

\subsection{Synthesis and Structure}\label{XRD}

Flux synthesis yielded  ZnSiP$_2$ crystals which are translucent, dark red in color, and have dimensions up to 2 mm thick, 5 mm wide, and 20 mm long (see Fig.~\ref{crystal}~(a)). Single crystal X-ray diffraction (SCXRD) data confirms that ZnSiP$_2$ crystallizes in the I$\bar{4}$2d space group (No. 122) with unit cell dimensions of \mbox{$a$ = 5.3986(2)} and \mbox{$c$ = 10.4502(6)~\AA} (see Fig.~\ref{crystal}~(b)). The SCXRD structural analysis results are given in Supplementary Tables~\ref{ZnSiP2_crystal_props}-\ref{ZnSiP2_X} and agree well with previously reported structural data.\cite{abrahams1970}

Some chalcopyrites are known to display atomic disorder resulting in variation of their optical properties.\cite{scanlon2012, seryogin1999, dietz1994, shay1975} Refinements of the SCXRD data indicate no significant disorder of Zn and Si between the 4a (Zn) and 4b (Si) sites. When occupancies and atomic displacement parameters were allowed to refine, the occupancies of all sites remained stable at 1.00(2) (see Supplementary Table~\ref{ZnSiP2_atom_coords}), indicating that all sites are fully occupied. A structural model with Zn and Si occupying both 4a and 4b sites was also employed. This model led to less than 1.3\% occupancy of Si on the 4a site and no mixing of Zn onto the 4b site. Statistics for this model are comparable to statistics for the model with no site mixing. In addition, all atomic displacement parameters led to nearly spherical thermal ellipsoids. Consistent with previous results, and in contrast to other ternary and quarternary PV materials,\cite{zawadzki2015, rey2014, scanlon2012, stjean2010} ZnSiP$_2$ prefers a highly ordered atomic structure that is very stable.\cite{abrahams1970, pamplin1979, martinez2015}

\subsection{Photoluminescence Characterization}

\begin{figure*}[t]
\centering
  \includegraphics[width=17.1 cm]{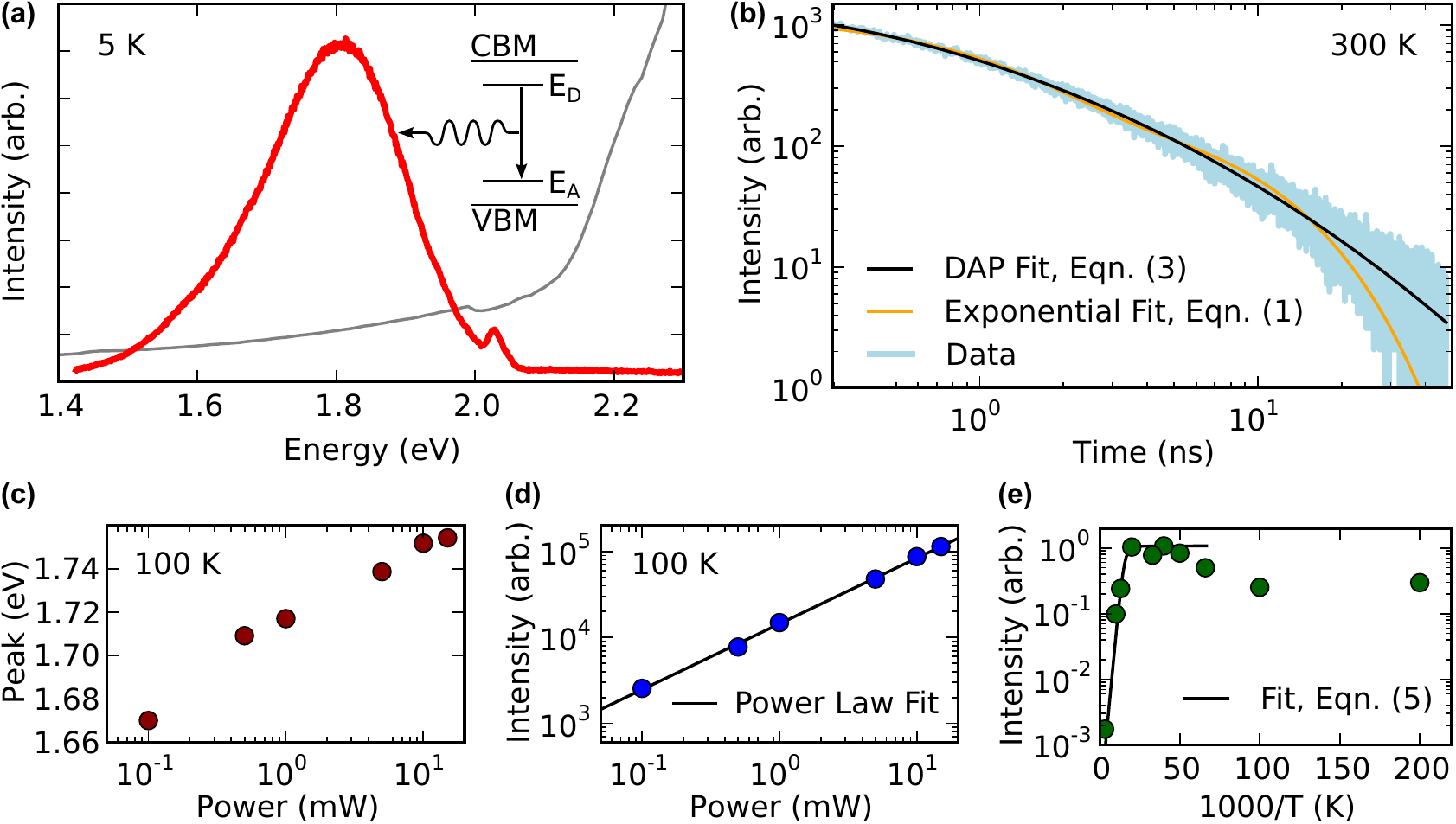}
  \caption{(a) Photoluminescence spectrum of ZnSiP$_2$ (red) and absorption edge (grey). The inset energy level diagram illustrates the most likely mechanism for the observed luminescence: donor/acceptor pair (DAP) transitions. (b) Time dependence of the 1.8~eV peak is evidence of DAP transitions and shows good carrier lifetime. (c) The power dependence of the 1.8~eV peak position shows a blue shift characteristic of DAP transitions. (d) The power dependence of the integrated intensity of the 1.8~eV peak; a fit to the data with $I(P)\propto~P^{k}$ further supports DAP transitions. (e) The temperature dependence of the integrated intensity of the 1.8~eV peak fits  Eqn.~(\ref{IofT}), giving $E_D=58$~meV.}
  \label{PL}
\end{figure*}

A combination of photoluminescence (PL) measurements and first-principles calculations (Section 2.4) of intrinsic point defects (vacancies, antisite defects) were used to understand the recombination mechanisms in ZnSiP$_2$.  Photoluminescence measurements were performed across a range of temperature (5--300~K) and excitation power (0.1--15~mW).  A large PL peak at 1.8~eV (Fig.~\ref{PL}~(a)) was observed at all measurement temperatures.  Consistent with prior work, a second small sharp peak at 2.04~eV was observed at temperatures below 15~K and is attributed to excitons.\cite{martinez2015} The time dependence of the 1.8~eV peak was measured to further probe the minority carrier recombination rates.  In concert, these measurements suggest donor-acceptor pair (DAP) recombination via shallow defects. 

Figure~\ref{PL}~(b) shows the time dependence of the PL intensity associated with the 1.8 eV, at room temperature.  An approximation of the carrier lifetime can be achieved using a biexponential model of the form

\begin{equation}
I(t) = A \, e^{-t/\tau_s}+B \, e^{-t/\tau_l},
\label{biexponential}
\end{equation}

\noindent
where $A$, $B$, $\tau_s$, and $\tau_l$ are fit parameters (orange curve in Figure~\ref{PL}~(b)).  This model assumes there are two independent, dominant decay mechanisms, and that each behaves exponentially. The parameters $\tau_s$ and $\tau_l$ are short and long lifetimes, respectively, and were found to be $\tau_s=0.9$~ns and $\tau_l=7.1$~ns.  This lifetime is comparable with those of many well developed thin film photovoltaic materials such as CdTe ($\sim$20~ns) and Cu$_2$ZnSnSe$_4$ ($\sim$18~ns).\cite{ma2013, guo2014}  From prior work, the hole mobility, $\mu_p$, has been measured as high as 25~cm$^2$/Vs.\cite{siegel1976, averkieva1983} For these n-type crystals, a hole diffusion length (at the surface) of 680~nm is calculated from $\mu_p$ and $\tau_l$. 

If DAP transitions are the primary PL mechanism, the recombination kinetics  depend on the wavefunction overlap between the spatially separated electron, trapped on a donor site, and hole, trapped on an acceptor site.\cite{thomas1965, yu2010} As such, the PL recombination is more complex than the two-process model assumed above.\cite{thomas1965}  The recombination rate for DAP transitions as a function of the distance, $r$, between donor and acceptor is assumed to be of the form

\begin{equation}
W(r) = W_0 \, exp\left[ -2 \, r/a\right],
\label{rate}
\end{equation}

\noindent
where $W_0$ is the rate as $r \rightarrow 0$ and $a$ is the larger of the donor or acceptor effective Bohr radii.\cite{thomas1965}  The PL intensity as a function of time, $t$, is proportional to

\begin{align}
I(t) &\propto exp\left[4 \, \pi \, N \!\! \int_{0}^{\infty} (exp\left[-W(r) \, t\right]-1) \, r^2 \, dr \right] \nonumber \\
&\qquad {} \times 4 \, \pi \, N \!\! \int_{0}^{\infty} W(r) \, exp\left[-W(r) \, t\right] \, r^2 \, dr,
\label{I_t}
\end{align}

\noindent
where $N$ is the concentration of the majority defect (possibly donors in this case because PEC rectification indicates \emph{n}-type conductivity).\cite{thomas1965}   While $W_0$, $a$, and $N$ are free parameters, the parameters $a$ and $N$ are coupled into the non-dimensional constant $\zeta \equiv a^3N$, and Eqn.~(\ref{rate})~and~(\ref{I_t}) were fit to the data in non-dimensional form (see Eqn.~(\ref{non_dim_rate})~and~(\ref{non_dim_I_t})). The black curve in Figure~\ref{PL}~(b) shows that the resulting fit is significantly better than the biexponential fit (orange), despite fewer free parameters. Because the  donor and acceptor concentrations are similar (as suggested by our theoretical analysis in section~\ref{defects}), the donor and acceptor contributions  to $\zeta = (2.3\pm0.3)  \times 10^{-2}$ cannot be decoupled.  $W_0$ was found to be $8 \times 10^9$~s$^{-1}$, slightly higher than GaAs ($3 \times 10^8$~s$^{-1}$).\cite{dean1973}  The fit of Eqn.~(\ref{I_t}) to the time dependent data (Fig.~\ref{PL}~(b)) demonstrates that DAP transitions are likely the dominant recombination mechanism.  This conclusion is in contrast to previous analysis of time dependent cathodoluminescence done at 10 K rather than room temperature, which determined the recombination to be due to bound excitons with phonon sidebands.\cite{shay1970} This difference is likely due to different recombination mechanisms dominating at different temperatures.

The power dependence of the PL gives additional evidence of DAP transitions. With increased injection level, the increased density of excited charge carriers leads to a stronger Coulomb interaction between ionized DAP's. This interaction leads to an increase in the emitted photon energy following:

\begin{equation}
\hbar \omega(r) = E_g - E_A - E_D + \frac{e^2}{r} (cgs),
\label{E_DAP}
\end{equation}

\noindent 
where $E_g$ is the band gap, $E_A$ is the ionization energy of the acceptor, $E_D$ is the ionization energy of the donor, $e$ is the electron charge, and $r$ is the distance between the donor and acceptor responsible for the transition. Figure~\ref{PL}~(c) shows the peak position shifting to higher energy as the injection level is increased, characteristic of DAP transitions.

Finally, the power dependence of the peak intensity (Fig.~\ref{PL}~(d)) can be considered within the DAP hypothesis. An exponential model of the form  $I(P)\propto~P^{k}$, where $P$ is the excitation power and $k$ is the exponent, was fit to the experimental data.  For the nine distinct temperatures at which power dependent data was collected, the average $k$ was 0.72 with a standard deviation of the average of 0.12. With \mbox{$k<1$}, the power dependence of the peak intensity is characteristic of either free to bound, or DAP transitions as being the primary source of radiative recombination.\cite{schmidt1992}

In addition to recombination mechanisms, PL data analysis can provide the donor activation energy (the material is \emph{n}-type, as shown in Sec.~\ref{PEC}).  Fig.~\ref{PL} (e) shows that the integrated intensity was found to increase significantly with decreasing temperature down to 25~K.  Below 25~K, the PL intensity decreased and eventually appeared to saturate.  The temperature region above 25~K was fit with

\begin{equation}
I(T)\propto\frac{1}{1+\alpha\,e^{(E_D/k_B T)}},
\label{IofT}
\end{equation}

\noindent
where $\alpha$ is a process rate parameter, $k_B$ is Boltzmann's constant, and $T$ is the temperature.\cite{krustok1997} Six temperature-dependent data sets were collected at different excitation powers, and the activation energy of the 1.8~eV peak was found to be \mbox{$E_D=58\pm16$~meV}; this value is shallow, and in agreement with previous reports.\cite{nahory1970, shay1970}

Both the time and power dependence of the PL intensity and the power dependence of the peak position indicate DAP recombination.  There was no evidence in the PL of deep level defects that would be detrimental to device operation, and the temperature dependence of the PL intensity predicts the donor level to be shallow, corresponding with the high $V_\text{oc}$ found in the PEC characterization.

\subsection{Identification of Potential Intrinsic Defects}\label{defects}

\begin{figure}
\centering
  \includegraphics[width=17.1cm]{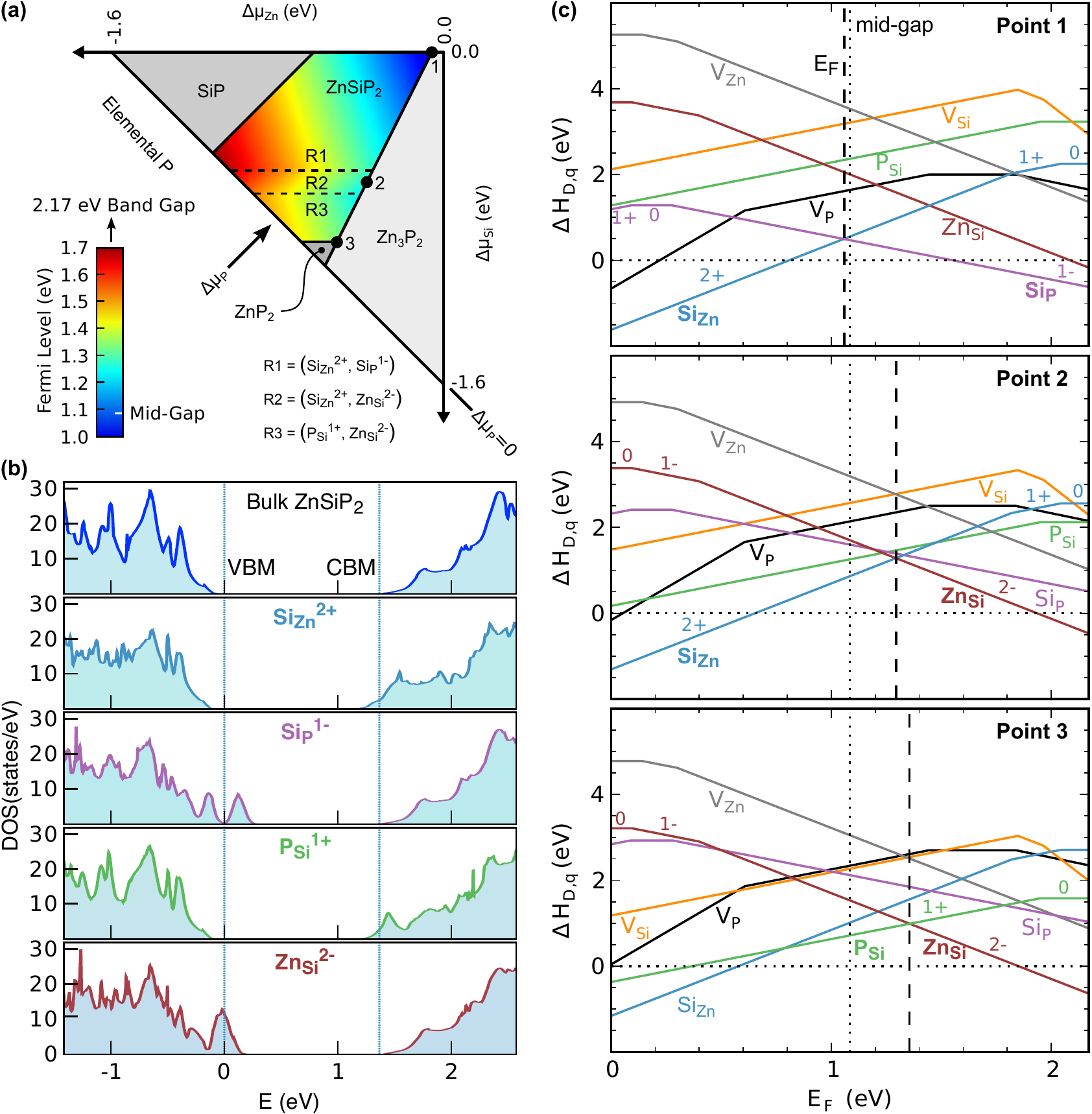}
  \caption{(a) Zn-Si-P phase diagram in chemical potential phase space. The flux growth conditions are Zn rich and are therefore expected to lie in the region between point 1 and point 3. Also shown is a heat map representing the Fermi level (E$_F$), which ranges from near mid-gap at point 1 to $\sim$1.7~eV above the valence band maximum. R1, R2 and R3 are regions in the chemical potential phase space where different donor-acceptor pairs are dominant. (b) Defect formation enthalpies ($\Delta$H$_{D,q}$) of 7 different point defects (vacancies and antisites) in all possible charge states (q ranging from 4- to 4+) at points 1, 2, and 3 on the phase diagram (a). The Fermi level is shown as a vertical dashed line along with mid-gap which is shown as a vertical dotted line; the relevant defects which set the Fermi level are labeled in bold. For additional details on all charge states of each point defect see Fig.~\ref{Defects_Flux_ESI}. (c) The density of states (DOS) of ZnSiP$_2$, without defects, and ZnSiP$_2$ with four key native point defects (Si$_{\text{Zn}}^{2+}$, Si$_{\text{P}}^{1-}$, P$_{\text{Si}}^{1+}$, and Zn$_{\text{Si}}^{2-}$). The associated defect states are shallow and appear as shoulders of the relevant the band edges; there is no evidence of mid-gap states that could trap carriers and detrimentally affect $V_\text{oc}$.}
  \label{Defects_Flux}
\end{figure}

To further understand charge carrier recombination in ZnSiP$_2$, first-principles density functional theory (DFT) calculations were employed to identify possible point defects that act as donors and acceptors in ZnSiP$_2$. By calculating the enthalpies of formation of known Zn-Si-P phases, a stability map in chemical potential space was constructed (Fig.~\ref{Defects_Flux}~(a)).\cite{stevanovic2012} The enthalpy of formation of intrinsic defects within the ZnSiP$_2$ region of stability of this chemical potential space were also calculated (Fig.~\ref{Defects_Flux}~(b)), along with the associated density of states (DOS) of relevant defects (Fig.~\ref{Defects_Flux}~(c)). 
The formation enthalpies ($\Delta$H$_\mathrm{D,q}$) of 7 different point defect types (3 vacancies, 4 antisite defects) were calculated in 7 different charge states (3- through 3+; 4- and 4+ were additionally examined for V$_\mathrm{Si}$) as a function of Zn, Si and P chemical potentials, and Fermi level (E$_\mathrm{F}$). Although previous electron paramagnetic resonance experiments have shown evidence of Zn and P vacancies, as well as Si on Zn antisite defects,\cite{gehlhoff2003, gehlhoff2003_2} these are the first calculations identifying likely point defects as a function of chemical potential for ZnSiP$_2$.

The Zn-Si-P phase diagram calculated at standard conditions using Fitted Elemental-Phase Reference Energies (FERE) is shown in Fig.~\ref{Defects_Flux}~(a).\cite{stevanovic2012} ZnSiP$_2$ occupies a large region of this chemical potential phase space, suggesting a large window for synthesis conditions. Competing phases include Zn, Si and P elements, and the binary compounds SiP, Zn$_3$P$_2$, and ZnP$_2$. Within the single phase region, where ZnSiP$_2$ is stable, there are three distinct regimes (labeled R1, R2, and R3) with different lowest $\delta H_{D,q}$ defect pairs. The dominant DAP's that set the Fermi level are Si$_{\text{Zn}}^{2+}$ and Si$_{\text{P}}^{1-}$ in Region 1, P$_{\text{Si}}^{1+}$ and Zn$_{\text{Si}}^{2-}$ in Region 2, and Si$_{\text{Zn}}^{2+}$ and Zn$_{\text{Si}}^{2-}$ in Region 3. The Fermi levels ($E_F$) are indicated by dashed vertical lines in Fig.~\ref{Defects_Flux}~(b).

The ZnSiP$_2$ crystals were grown under Zn-rich, Si-poor, and P-poor conditions. These growth conditions most likely lie near the line connecting points 1, 2, and 3, shown in Fig.~\ref{Defects_Flux}~(a).
For this reason, the defect formation enthalpies and their concentrations were examined at these points which lie in the corresponding regions R1, R2, and R3, in Fig.~\ref{Defects_Flux}~(a). 
For the point defects considered, $\Delta H_\mathrm{D,q}$ as a function of $E_\mathrm{F}$ is shown in Fig.~\ref{Defects_Flux}~(b) wherein the formation enthalpy of the lowest-energy DAP at each of the three points is shown in bold.
A first order approximation of the defect concentrations can be determined from

\begin{equation}
\left[C\right]=\left[S\right]\,e^{-\frac{\Delta H_\mathrm{D,q}}{k_B \, T}},
\label{defect_conc}
\end{equation}

\noindent
where $\left[S\right]$ is the crystallographic site concentration ($\left[\text{Zn}\right]=\left[\text{Si}\right]=\left[\text{P}\right]/2\approx1.3\times10^{22}$~cm$^{-3}$), $k_B$ is the Boltzmann constant, $T$ is the temperature at which the defects form, and $\Delta H_\mathrm{D,q}$ is evaluated at the chemical potentials associated with a particular point in Fig.~\ref{Defects_Flux}~(a) and the 

\onecolumn

\newpage

\twocolumn

\newpage

\noindent corresponding Fermi level. The defect concentrations were calculated at $T=500\,^\circ$C, resulting in a lower limit to what may be observed experimentally; this is because the crystal growth occurs as the temperature of the Zn flux slowly cools ($\sim$2~$^\circ$C/h) from 1000~$^\circ$C to 500~$^\circ$C, after which the flux is rapidly cooled to room temperature. The calculated DAP's and their associated concentrations are given in Table \ref{ZnSiP2_DAPs}.

\begin{table}
\caption{\small Concentrations of dominant point defects in ZnSiP$_2$ at 500~$^\circ$C. The locations of points 1, 2, and 3 are identified in Fig.~\ref{Defects_Flux}~(a). $\left[D\right]$ and $\left[A\right]$ are the donor and acceptor concentrations in cm$^{-3}$.}
\begin{center}
\begin{tabular} {l  c  c  c} 
\hline\\ [-1.5ex]
Point		&DAP 	&$\left[D\right]$			&$\left[A\right]$\\ [1ex]
\hline\\ [-1.5ex]
1			&Si$_{\text{Zn}}^{2+}$/Si$_{\text{P}}^{1-}$			&$7.5\times10^{18}$			&$1.5\times10^{19}$\\ [1ex]
2			&Si$_{\text{Zn}}^{2+}$/Zn$_{\text{Si}}^{2-}$			&$6.1\times10^{13}$			&$6.1\times10^{13}$\\[1ex]
3			&P$_{\text{Si}}^{1+}$/Zn$_{\text{Si}}^{2-}$			&$4.3\times10^{15}$			&$4.3\times10^{15}$\\[1ex]
\hline
\end{tabular}
\end{center}
\label{ZnSiP2_DAPs}
\end{table}

The density of states (DOS) of defect-free ZnSiP$_2$ and ZnSiP$_2$ containing one of the four prominent defects are shown in Fig~\ref{Defects_Flux}~(c). The shallow donor and acceptor states appear as shoulders near the conduction band minimum (CBM) and the valence band maximum (VBM), respectively. There is no evidence of mid-gap states in any case. Previous studies have observed a deep acceptor level, $\sim$0.7~eV above the VBM, which appears to have been the result of extrinsic impurity (Cu) doping.\cite{ziegler1973, gorban1976, endo1992}

PL experiments and theoretical defect calculations are consistent with DAP antisite defects as the prevalent recombination mechanism in ZnSiP$_2$.
These DAP defects work together to set the Fermi level such that ZnSiP$_2$ is expected to be intrinsic, or moderately \emph{n}-type, across a wide range of synthesis conditions. Analysis of the temperature dependence of the PL intensity found the donor level to be shallow: $\sim$58~meV below the CBM.  Correspondingly, the dominant defects identified in the calculations were found to occupy shallow levels, based on their associated DOS.  These results indicate that, over the full chemical potential region of stability, ZnSiP$_2$ PV devices are defect-tolerant and should have a high $V_\text{oc}$, in agreement with the PEC results from Section \ref{PEC}.

\begin{figure}
\centering
  \includegraphics[width=\columnwidth]{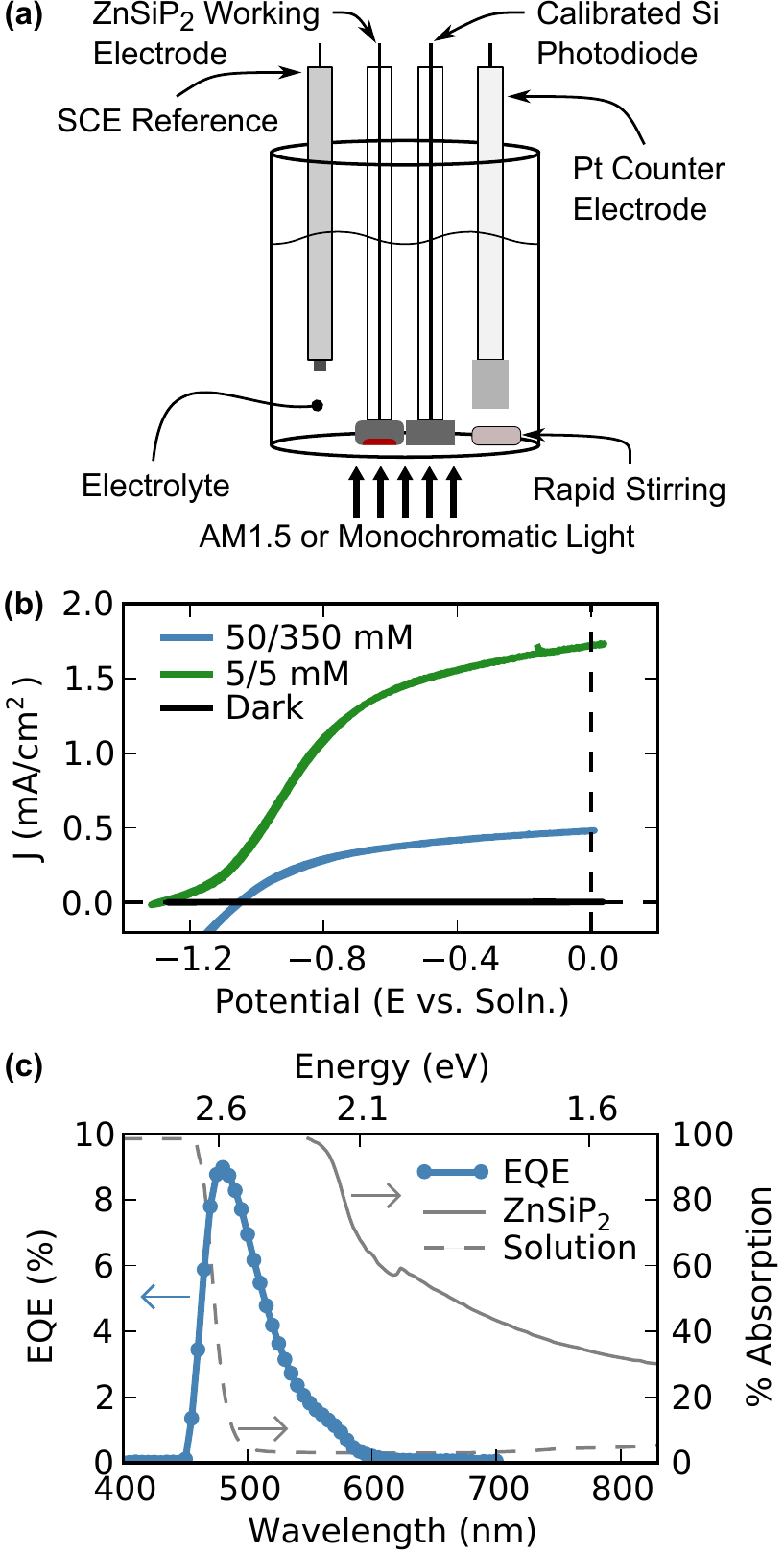}
  \caption{(a) Schematic of the photoelectrochemical cell with a ZnSiP$_2$ working electrode and Si photodiode for calibrating light intensity. (b) Photoelectrochemical $J-E$ data for ZnSiP$_2$ electrodes under 1 Sun illumination in contact with 50/350 mM Fe(CN)$_6^{(3^-/4^-)}$ (blue), and 5/5 mM Fe(CN)$_6^{(3^-/4^-)}$ (light=green, dark=black),  (c) Spectral response of the same (blue in (b)) ZnSiP$_2$ electrode in 50/350 mM solution, along with the percent absorption of 3~mm of solution and of the 0.5~mm thick ZnSiP$_2$ electrode.}
  \label{EChem}
\end{figure}

\subsection{Photovoltaic Characterization}\label{PEC}

Regenerative photoelectrochemistry (PEC) was used to characterize the photovoltaic properties of ZnSiP$_2$. This technique allows the measurement of the current density vs. potential ($J$-$E$) of irregularly shaped bulk single crystals without the need to create a solid state \emph{p}-\emph{n} junction. A ZnSiP$_2$ crystal is formed into an electrode and submersed in an electrolyte solution containing a one-electron, reversible, outer-sphere redox couple, creating a semiconductor-liquid junction (Fig~\ref{EChem}~(a)). This method has proven to be a successful initial characterization technique for many semiconductors.\cite{warren2012, santori2012, boettcher2010, tufts1990, rosenbluth1986, heller1980, hodes1976} Within this photoelectrochemical configuration, initial measurements as high as 0.9\% photoconversion efficiency were obtained from ZnSiP$_2$ single crystals in contact with the Fe(CN)$_6^{(3^-/4^-)}$~(aq.) redox couple.

Photoelectrochemical electrodes were constructed from several single crystals, and representative $J$-$E$ data is shown in Fig.~\ref{EChem}~(b). The representative $J$-$E$ data was collected for electrodes in contact with 50~mM K$_3$Fe(CN)$_6$/350~mM K$_4$Fe(CN)$_6$ (aq) and 5~mM K$_3$Fe(CN)$_6$/5~mM K$_4$Fe(CN)$_6$ (aq).\cite{strandwitz2013} The photoelectrochemical rectification with the Fe(CN)$_6^{(3^-/4^-)}$~(aq) electrolyte indicates that the ZnSiP$_2$ crystals are \emph{n}-type. 

In the 50/350 mM Fe(CN)$_6^{3-/4-}$ electrolyte, an $E_\text{oc}$ of 1.08~V,  a short circuit current density ($J_\text{sc}$) of 0.73~mA/cm$^2$, and a fill-factor ($FF$) of 0.53 were measured.  However, at these concentrations, the redox couple effectively absorbs all incident AM1.5 light at energies above 2.6 eV, which is a significant fraction of the illumination above ZnSiP$_2$'s 2.1eV optical band gap (see Fig~\ref{EChem}~(c)). With a reduced concentration of 5/5 mM Fe(CN)$_6^{3-/4-}$, better $E_\text{oc}$  and $J_\text{sc}$ were obtained due to reduced absorption losses. At this lower concentration of the redox couple the champion electrode produced an $E_\text{oc}$ of 1.30~V, a $J_\text{sc}$ of 1.72~mA/cm$^2$, a $FF$ of 0.41, and an efficiency of 0.9\%. This unoptimized cell configuration is limited by solution resistance, contact resistance, and concentration overpotential, each of which could be significant in these preliminary electrochemical measurements. The $J$-$E$ data has not been corrected for mass transport limitations or concentration overpotential, and thus the low $FF$ does not reflect the intrinsic performance of the semiconductor.\cite{santori2012}

High $E_\text{oc}$ ($V_\text{oc}$ in a solid-state device) is indicative of good material quality, good charge carrier lifetime, and a lack of deep level defects. 
An ideal homojunction ZnSiP$_2$ PV device with optimized doping of the base and emitter would be expected to have a maximum $V_\text{oc}$ of $\sim$1.8~V. \cite{baruch1995, levy2006} The Fe(CN)$_6^{(3^-/4^-)}$ redox couple is not optimized for the band edge positions of ZnSiP$_2$, nor has the effect of doping of the ZnSiP$_2$ crystals been investigated.  Therefore, the observed 1.30~V $E_\text{oc}$ of ZnSiP$_2$ under 1 sun conditions is promising because increasing the $V_\text{oc}$ can be one of the most significant challenges in the development of a new material.\cite{wei1998} The high V$_{oc}$ and low band gap-voltage offset observed here are on par with much more mature wide band gap III-V materials in solid-state devices,\cite{Masuda2015, Lu2014, Cornfeld2012} but in a material that can be easily integrated with Si. For materials lattice matched to Si, much lower voltages are typically observed, even in more optimized solid-state devices. For example, GaNPAs devices on Si have a reported V$_{oc}$ of 1.1 V, as well as poor fill factor and current.\cite{Geisz2005} A leading contender for earth-abundant tandem PV on Si is CZTS, but these devices on Si have a reported V$_{oc}$ of only 0.12 V,\cite{Song2015} and optimized devices not grown on Si have achieved only 0.66~V.\cite{Shin2013} Thus, it is extremely promising that in this first report of photovoltaic performance of ZnSiP$_2$, we have observed such a high voltage.

There are several potential explanations for the observed low current density of the electrodes: diffusion limitations due to crystal thickness, weak light absorption, resistive losses, and parasitic light absorption by the solution. To better understand the photocurrent generation in these devices, the electrochemical spectral response was investigated. Figure~\ref{EChem}~(c) shows the external quantum efficiency (EQE) of a typical ZnSiP$_2$ electrode in 50/350mM Fe(CN)$_6^{3-/4-}$  (aq) along with the percent absorption by the solution (UV-Vis), and the percent absorption by the ZnSiP$_2$ (photothermal deflection spectroscopy\cite{martinez2015}). The $J$-$E$ performance of the same electrode under AM1.5G illumination is shown (blue) in Fig~\ref{EChem}~(b). From the spectral response, it is apparent that the short wavelength spectral region ($>$~2.6~eV) is strongly absorbed by the electrolyte so only a small band of the AM1.5 spectrum ($\sim$2.1-2.6~eV) is absorbed by the crystal. The long wavelength edge of the spectral response closely resembles the ZnSiP$_2$ absorbance spectrum, indicating that photocurrent is limited by light absorption, as expected due to the symmetry forbidden nature of transitions at the band gap.\cite{alekperova1969, shay1973, martinez2015} This arises because the electric dipole transition between the valence band maximum (VBM) and conduction band minimum (CBM) is forbidden by a selection rule.\cite{yu2010} The crystals were also thick (0.5-2 mm) relative to the expected absorption depth of the photons, so it is likely that many photocarriers recombined in the bulk before they could be collected. The total $J_\text{sc}$, as calculated from integrating the spectral response, was 0.35~mA/cm$^2$, 27\% lower than that measured in the $J$-$E$  characteristic (Fig.~\ref{EChem}~(b)).  While the spectral response and $J$-$E$ characteristic were collected from the same ZnSiP$_2$ electrode, the experimental setups were different in each case.  These differences are primarily responsible for the discrepancy in $J_\text{sc}$ between the two different measurements.

\section{Conclusions}

A combination of photoeletrochemical (PEC) characterization, photoluminescence (PL) measurements, and theoretical intrinsic defect calculations have shown that ZnSiP$_2$ has the potential to be a high quality photovoltaic top cell material. PEC characterization revealed an open circuit voltage of 1.3~eV, indicating a lack of deep level defects.  PL confirmed the presence of shallow donor-acceptor pairs as well as the lack of deep level defects. Theoretical defect calculations established that the dominant defects in ZnSiP$_2$ are shallow antisite defects throughout the stability window in Zn-Si-P chemical potential space.  ZnSiP$_2$ is additionally predicted to be intrinsic, or moderately \emph{n}-type throughout this entire region in chemical potential space. Thus, this material is intrinsically very stable and defect-tolerant. In this first report of the solar energy conversion properties of ZnSiP$_2$, we have shown that it has excellent potential as a top cell on Si due to its abundance, defect tolerance, high voltage potential, and wide phase stability window. 

\section{Methods}
\subsection{Experimental}

Large single crystals of ZnSiP$_2$ (needle shaped or sometimes hollow tubes, up to 2 by 5 by 20 mm) were synthesized using a zinc self-flux.\cite{martinez2015PVSC} Zn (Alfa Aesar, \#10759, 99.999\% metals basis), Si (Alfa Aesar, \#43006, 99.9999\% metals basis) and P (Alfa Aesar, \#10670, 99.999\% metals basis) were mixed in 20:1:2.5 molar ratio and sealed in a 19~mm inner diameter quartz ampule under static vacuum better than 1e-6~mbar. To facilitate dissolution in the flux, the silicon was ball milled (Spex 8000 Mixer/Mill) in silicon carbide vials to a fine powder before use. The sealed ampule was placed vertically in a tube furnace and heated to 1000 $^{\circ}$C and soaked for 30 hrs to ensure complete dissolution and homogenization. After the soak the furnace was cooled at 2 $^{\circ}$C/hr to 500 $^{\circ}$C when it was removed. The excess molten flux was immediately decanted by tilting the hot ampoule before it cooled to room temperature. Residual Zn flux was removed from the crystals by etching with hydrochloric acid (37\% by volume) mixed with deionized water in 1:3 ratio by volume. Residual Si precipitates were removed from the crystal surfaces by etching in a dilute mixture of hydrofluoric (49\% by volume) and nitric (70\% by volume) acids in deionized water in 10:3:87 volume ratio. Other precipitates or inclusions have not been observed. The crystals did not receive further treatment before measurements were performed except when otherwise noted.

A fragment (approximately 0.01$\times$0.01$\times$0.03~mm) cut from a larger single crystal was used to collect single crystal X-ray diffraction (SCXRD) data. This crystal fragment was mounted onto the goniometer of a Bruker KappaCCD diffractometer equipped with Mo~K$_{\alpha}$ ($\lambda$ = 0.71073~\AA) radiation. Data collection and structure solution were performed with SHELX-97. Structural refinements and extinction corrections were performed using SHELXL suite.

The photoelectrochemical (PEC) response of ZnSiP$_2$ was tested using the Fe(CN)$_6^{(3-/4-)}$ redox couple (potassium ferrocyanide and potassium ferricyanide in H$_2$O with no supporting electrolyte).\cite{scheuermann2013, strandwitz2013} ZnSiP$_2$ electrodes were constructed by evaporating metal back contacts on one flat side of the crystals.  The back contacts were 700~nm of In, and they were capacitively blasted to achieve Ohmic behavior. A copper lead wire was affixed to the back contact with silver paint and the electrode was sealed at the end of a glass tube with epoxy (Loctite, Hysol 1C). An AM1.5G light source (ABET Technologies, model no. 10500) was used to illuminate the front of the crystal and the light intensity was set by placing a calibrated photodiode (Thor Labs part no. FDS100) electrode in the same position as the crystal electrode, without electrolyte present, resulting in very repeatable data.  Current-voltage data was collected using a BioLogic SP-240 potentiostat with Pt foil as the counter electrode and a saturated calomel reference electrode.  Potentials were later adjusted to reference the solution potential, $E(A/A^-)$, which was measured with a Pt wire in the electrolyte.  

Non-aqueous PEC data was taken using the same electrodes, which were briefly etched in dilute HCl, and then inserted into a 3-neck PEC cell on a schlenk line.  The electrolyte consisted of 10~mM ferrocene (Fc, Sigma-Aldrich), and 0.05~mM ferrocenium dissolved in acetonitrile with 0.1~M LiClO$_4$ as a supporting electrolyte.\cite{Ritenour2012}  The same ABET model of light source was used for illumination and the photon flux was calibrated using a photodiode using a similar method to the aqueous tests.

Photoluminescence spectra were collected with a 442 nm HeCd laser with a 250~$\mu$m spot size.  Spectra were collected at temperatures of 5, 10, 15, 20, 25, 30, 50, 75, 100, and 300 K.  At each temperature, the laser power was varied and spectra were collected at 0.1, 0.5, 1.0, 5.0, 10.0, and 15.0 mW (laser power at crystal surface).

A Yb:KGW laser and an optical parametric amplifier (pulse length 0.3~ps, repetition rate 1.1~Mhz, wavelength 420~nm) were used for single photon excitation photoluminescence lifetime measurements. The excitation beam diameter was 0.2~mm and average laser power was 2~mW. Photoluminescence was measured with a 10~nm bandpass filter at 650~nm, a Si avalanche photodiode detector, and time-correlated single photon counting.  Measurements were done at room temperature on two different, opposing sides of the same crystal.  The results were slightly different from one side to the other and measurements of different spots on the same side yielded very consistent data.

\subsection{Computational}

To study the intrinsic point defect physics in ZnSiP$_2$, we performed density functional theory (DFT) calculations using the VASP code \cite{kressePRB1996}, in the generalized gradient approximation (GGA) of Perdew-Burke-Ernzerhof, \cite{Perdew1996} with the projector augmented wave formalism \cite{blochlPRB1994}.
The total energies were calculated for supercells containing 64 atoms (2$\times$2$\times$2 unit cells) using a plane-wave cutoff of 400 eV and a 4$\times$4$\times$4 Monkhorst Pack k-point grid. 
The defect supercells were relaxed following the numerical approach described in Ref. \citenum{stevanovic2012} for application of the electrostatic Hubbard U.
Defect supercell total energies were calculated for $D$ = V$_\text{Zn}$, V$_\text{Si}$, V$_\text{P}$, Zn$_\text{Si}$, Si$_\text{Zn}$, Si$_\text{P}$, P$_\text{Si}$ in charge states $q=$ -3, -2, -1, 0, 1, 2, 3 for all defects and also $q=$ -4, 4 for V$_\text{Si}$.
Using the calculated total energies, the defect formation enthalpies ($\Delta H_{D,q}$) are calculated as

\begin{equation}
\Delta H_{D,q} = (E_{D,q} - E_H) + \sum\mu_{\alpha} + qE_F
\label{defectform}
\end{equation}

\noindent
where the first term on right hand side is the energy difference between the supercell with defect ($D$) in charge state $q$ and the defect-free host ($E_H$).  The second and the third terms describe the chemical potential of the atomic and electronic reservoirs, respectively. The chemical potential $\mu_\alpha$ ($\alpha=$ Zn, Si, P) is given, relative to the elemental phase, such that $\mu_\alpha = \mu_\alpha^0 + \Delta\mu_\alpha$, where  $\mu_\alpha^0$ is the elemental phase chemical potential, calculated using the Elemental Phase Reference Energies (FERE) \cite{stevanovic2012} and $\Delta\mu_\alpha$ is the deviation in chemical potential from the elemental reference state. The Fermi energy ($E_F$) is given with respect to the valence band maximum (VBM) i.e. $E_F = E_v + \Delta E_F$; the value of $E_F$ can vary from zero to the band gap. 

\balance

The underestimation of the band gap within the DFT-GGA method was ``corrected" by applying band edge shifts, which are based on the GW quasi-particle energy calculations, as described in Ref. \citenum{PengPRB2013} and \citenum{stevanovic2014}. The band edge shifts for ZnSiP$_2$ were calculated to be $\Delta E_{VBM}=$ -0.907 eV and $\Delta E_{CBM}=$ -0.084 eV. The band gap, after application of the GW-calculated shifts, is 2.17 eV, in good agreement with experimental measurements.\cite{martinez2015, shirakawa1981, prochukhan1977} While the general methodology for calculation of defect formation enthalpy from supercell calculations is well established, it must be noted that several corrections need to be applied to obtain a more accurate defect formation enthalpy. Following the methodology described in Ref.~\citenum{Lany2009}, we have applied the following corrections to $\Delta H_{D,q}$: (1) image charge correction for charged defects, (2) potential alignment corrections for charged defects, (3) band filling corrections for shallow donors/acceptors, and (4) band gap corrections for shallow donors/acceptors. Some of these corrections (3, 4) depend on whether the defect is a shallow or deep level; we performed extensive tests, including DFT calculations with hybrid functionals (HSE06) \cite{krukauJCP2006} to ascertain the nature of the defect levels. To calculate the Fermi level, as a first approximation, we used the crossing point between the donor and acceptor with the lowest formation enthalpies on a $\Delta H_{D,q}$ vs. E$_F$ plot shown in Fig~\ref{Defects_Flux}~(b). The defect concentrations $\left[C\right]$ were calculated by multiplying the total number of lattice sites $\left[S\right]$ that can accommodate a certain defect, by the Boltzmann factor (Eqn.~\ref{defect_conc}).


\section*{Acknowledgements}

The authors thank Anna Duda for depositing electrical contacts. Funding for this work was provided by the National Renewable Energy Laboratory through the Laboratory-Directed Research and Development program and by the National Science Foundation through the Renewable Energy Materials Research and Engineering Center at the Colorado School of Mines under NSF grant number DMR-0820518. RTM acknowledges NSF CAREER Award 1541230 for support of this work.  KAB is thankful for funding from the Danish Council for Independent Research (DFF), grant No. 4090-00071, and the DFF Sapere Aude program. The U.S. Government retains and the publisher, by accepting the article for publication, acknowledges that the U.S. Government retains a nonexclusive, paid up, irrevocable, worldwide license to publish or reproduce the published form of this work, or allow others to do so, for U.S. Government purposes.






\newpage

\onecolumn

\section*{Electronic Supplementary Information}

\beginsupplement

\subsection*{Summary of II-IV-V$_2$ properties}

\begin{tablehere}
\caption{\small Experimental electronic \& optical properties of II-IV-V$_2$ compounds. Forbidden transitions are shaded in red. All band gaps are for chalcopyrite phases. Experimentally determined chalcopyrite lattice parameters ($a$ \& $c$) are in \AA, band gaps (E$_g^{exp}$ from experiment and E$_g^{calc} $ from our GW calculations) in {eV}, mobilities ($\mu$) in {cm$^2$/V\,s} and free carrier concentrations {($n$ \& $p$)} in  {cm$^{-3}$}.}
\begin{center}
\begin{tabular} {l  l  c  c  c  c  c  c  c  c  l} 
\hline
ID & Material & $a$ & $c$ & $c/a$ & E$_g^{exp}$ (E$_g^{calc} $) & $\mu_e$ & $n$ & $\mu_h$ & $p$ & References\\
\hline
0 & Si & 5.43 & --- & --- & 1.12 & $<$ 1500 & 10$^{14}$-10$^{19}$ & $<$ 500 & 10$^{14}$-10$^{19}$ & \cite{sze2006}\\
\hline
1 & ZnSiP$_2$ & 5.40 & 10.44 & 1.93 & \cellcolor[rgb]{1,0.6,.6} 2.0-2.3 (2.09)     	& 50-1,000 		& 10$^{13}$-10$^{18}$	& 1-25    		& $\leq$10$^{17}$ 
& \cite{pamplin1979, prochukhan1977, springthorpe1968, springthorpe1970, shay1975, ziegler1973, siegel1974, gorban1976, siegel1976, averkieva1983, yao1986}\\
2 & ZnGeP$_2$ & 5.49 & 10.80 & 1.97 & \cellcolor[rgb]{1,0.6,.6} 1.8-2.3 (2.10)      	&---   				& 10$^{13}$-10$^{15}$	& 20        	& 10$^{10}$-10$^{17}$ 
& \cite{goodman1957,springthorpe1968, prochukhan1977, pamplin1979, shay1975, yao1986}\\
3 & ZnSnP$_2$ & 5.65 & 11.30 & 2 & 1.6-2.1 (1.86)                                  		&---         			&---                       		& 55       	& 10$^{16}$-10$^{17}$ 
& \cite{goodman1957,prochukhan1977, pamplin1979, stjean2010, shay1975, shin1989}\\
4 & MgSiP$_2$ & 5.72 & 10.11 & 1.77 & \cellcolor[rgb]{1,0.6,.6} 2.2 (2.25)      				&---            	&---  					&---   		&--- 
& \cite{springthorpe1969, averkieva1983}\\
5 & CdSiP$_2$ & 5.68 & 10.44 & 1.84 & \cellcolor[rgb]{1,0.6,.6} 2.2 (2.10)        		& 200-1,000       			& 10$^{10}$-10$^{15}$ 	&---      	&--- 
& \cite{springthorpe1968, prochukhan1977, pamplin1979, shay1975, averkieva1983}\\
6 & CdGeP$_2$ & 5.77 & 10.82 & 1.88 & 1.6-1.8 (1.91)                                      	& 100         			& 10$^{11}$-10$^{14}$  	& 25           	& 10$^{10}$-10$^{15}$  
& \cite{goodman1957,springthorpe1968, prochukhan1977, pamplin1979, shay1975}\\
7 & CdSnP$_2$ & 5.90 & 11.52 & 1.95 & 1.0-1.5  (1.33)                                     	& 2,000     			& 10$^{15}$-10$^{18}$	&---       		& 10$^{14}$
& \cite{goodman1957,springthorpe1968, prochukhan1977, pamplin1979, shay1975}\\
\hline
8 & ZnSiAs$_2$ & 5.61 & 10.88 & 1.94 &  \cellcolor[rgb]{1,0.6,.6} 1.7-2.1 (1.53) 	& 40            			& 10$^{8}$                   	&140-170 	& 10$^{13}$-10$^{17}$ 
& \cite{goodman1957,springthorpe1968, prochukhan1977, pamplin1979, shay1975, wen1997,averkieva1983}\\
9 & ZnGeAs$_2$ & 5.67 & 11.15 & 1.97 & 0.6-1.1 (1.21)                              		&---          			&---                      		& 55         	& 10$^{16}$-10$^{19}$ 
& \cite{goodman1957,prochukhan1977, pamplin1979, peshek2013}\\
10 & ZnSnAs$_2$ & 5.85 & 11.70 & 2.00 & 0.6-0.7 (0.89)                           		&---        			& 10$^{15}$          		& 300          	& 10$^{17}$-10$^{21}$ 
& \cite{pamplin1979, shay1975, prochukhan1977}\\
11 & MgGeAs$_2$  & 5.66 &--- &--- &--- (1.82)                                                        		& 600      			& 10$^{18}$          		& 35           	& 10$^{19}$ 
& \cite{li2007}\\
12 & CdSiAs$_2$ & 5.88 & 10.88 & 1.85 & 1.5-1.6 (1.57)                                      	&---         			& 10$^{17}$                  	& 500        	& 10$^{14}$-10$^{17}$ 
& \cite{springthorpe1968, prochukhan1977, pamplin1979, shay1975,averkieva1983}\\
13 & CdGeAs$_2$ & 5.94 &11.22 & 1.89 & 0.5-0.6 (0.70)                                    	& 2,500    			& 10$^{16}$-10$^{18}$	& 1,500      	& 10$^{16}$-10$^{18}$
& \cite{pamplin1979, prochukhan1977}\\
14 & CdSnAs$_2$ & 6.09 & 11.94 & 1.96 & 0.3 (0.38)                               			& 12,000     			& 10$^{17}$-10$^{18}$ 	& 190           	& 10$^{17}$-10$^{18}$
& \cite{pamplin1979, shay1975, strauss1961, prochukhan1977}\\
\hline
\end{tabular}
\end{center}
\label{electronic_optical}
\end{tablehere}

\subsection*{ZnSiP$_2$ Single Crystal XRD Results}

\begin{tablehere}
\caption{\small Atomic coordinates and site occupancies for ZnSiP$_2$.}
\begin{center}
\begin{tabular} {l  c  c  c  c  c  c} 
\hline
Atom		&Wyckoff site 	&x					&y			&z			&Occupancy	&$U_{eq}$ (\AA$^2$)\\
\hline
Zn			&4a			&1/2				&0			&3/4		&0.99(2)			&0.005(1)\\
P			&8d			&0.73023(6)	&1/4		&1/8		&1.00(2)			&0.005(1)\\
Si			&4b			&0					&0			&1/2		&1.00(2)			&0.005(1)\\
\hline
\end{tabular}
\end{center}
\label{ZnSiP2_atom_coords}
\end{tablehere}

\newpage

\begin{tablehere}
\caption{\small Crystallographic data for ZnSiP$_2$.}
\begin{center}
\begin{threeparttable}
\begin{tabular} {l  c} 
\hline
Item [units]										&Value\\
\hline
Formula											&ZnSiP$_2$\\
Space group										&I$\bar{4}$2d (No.122)\\
Crystal system										&Tetragonal\\
a [\AA]											&5.3986(2)\\
c [\AA]											&10.4502(6)\\
V [\AA$^3$]										&304.57(2)\\
Z												&4\\
FW [g/mol]				&155.413\\
$\rho_{calcd}$ [g/cm$^3$]	&3.389\\
Atomic density [atoms/cm$^3$]		&$5.25\times10^{22}$\\
T [K]												&293(2)\\
$\lambda$ [\AA]									&0.71073\\
$\theta_{maximum}$									&49.81\\
Number unique reflections ($n$)						&798\\
Number reflections I $>$2$\sigma$(I)					&726\\
Number of refined parameters	 ($p$)					&11\\
Extinction coefficient									&0.046(2)\\
$\mu$ [mm$^{-1}$]									&9.18\\
Flack parameter									&0.023(6)\\
$R(int)$ [\%]										&1.93\\
$R (F)$\tnote{a} [\%]									&1.71\\
$Rw(F_o^2)$\tnote{b} [\%]							&3.79\\
$GOF(F^2)$\tnote{c}									&1.099\\
$\Delta\rho_{min}$, $\Delta\rho_{max}$					&-1.198, 0.583\\
\hline
\end{tabular}
\begin{tablenotes}
\item[a] $R(F) = \Sigma\left|\left|F_o\right| - \left|F_c\right|
\right|/\Sigma\left|F_o\right|$\\
\item[b] $Rw(F_o^2) = \left[\Sigma w(F_o^2 - F_c^2)^2/\Sigma w(F_o^2)^2\right]^{1/2}$\\
\item[c] $GOF(F^2) = \left[(\Sigma\left|w/\left|F_o^2-F_c^2\right|^2\right|)/(n-p)\right]^{1/2}$
\end{tablenotes}
\end{threeparttable}
\end{center}
\label{ZnSiP2_crystal_props}
\end{tablehere}

\begin{tablehere}
\caption{\small Anisotropic displacement parameters for ZnSiP$_2$.}
\begin{center}
\begin{tabular} {l  c  c  c  c  c  c} 
\hline
Atom		&$U_{11}$		&$U_{22}$		&$U_{33}$\\
\hline
Zn			&0.00463(5)	&0.00463(5)	&0.00522(6)\\
P			&0.00481(9)	&0.00455(8)	&0.0045(10)\\
Si			&0.0046(1)		&0.0046(1)		&0.0042(1)\\
\hline
\end{tabular}
\end{center}
\label{ZnSiP2_U}
\end{tablehere}

\begin{tablehere}
\caption{\small Selected bond distances (\AA) and angles ($^{\circ}$) for ZnSiP$_2$.}
\begin{center}
\begin{tabular} {l  c  l  c} 
\hline
Configuration		&Distance 		&Configuration		&Angle\\
\hline
Zn-P ($\times$4)			&2.3767(2)		&P-Zn-P ($\times$2)		&107.582(3)\\
								&					&P-Zn-P ($\times$2)		&113.320(7)\\
Si-P ($\times$4)			&2.2522(2)		&P-Si-P ($\times$2)		&109.102(7)\\
								&					&P-Si-P ($\times$2)		&109.656(4)\\
\hline
\end{tabular}
\end{center}
\label{ZnSiP2_X}
\end{tablehere}

In the SiP$_4$ tetrahedra, P-Si-P angles measure 109.102(7)$^{\circ}$ and 109.656(4)$^{\circ}$, which are close to the ideal 109.5$^{\circ}$ angles expected for tetrahedral geometry. In ZnP$_4$, however, two of the angles measure 107.582(3)$^{\circ}$ while two angles measure 113.320(7)$^{\circ}$, which indicate a distorted tetrahedral Zn environment.

\subsection*{Non-Dimensional Form of DAP Time Dependent Equations}

The non-dimensional variable, $x\equiv r^3\,N$, and parameter, $\zeta\equiv a^3\,N$, are defined from $r$, the distance between DAP's, $a$, the characteristic distance (attributed to the larger of the donor or acceptor effective Bohr radii\cite{thomas1965}), and $N$, the concentration of the majority defect. The recombination rate as a function of $x$ is then

\begin{equation}
W(x) = W_0 \, exp\left[ -2 \, \left(x/\zeta\right)^{1/3}\right],
\label{non_dim_rate}
\end{equation}

\noindent
where $W_0$ is the rate as $r \rightarrow 0$, as it is in Eqn.~\ref{rate}.  Given that $x=r^3\,N \rightarrow dx=3\,r^2\,N\,dr \rightarrow r^2\,dr=\frac{dx}{3\,N}$, the PL intensity as a function of time, $t$, according to Eqn~(\ref{I_t}), but in terms of $x$ instead of $r$, is proportional to

\begin{equation}
I(t) \propto exp\left[\frac{4}{3} \, \pi \!\! \int_{0}^{\infty} (exp\left[-W(x) \, t\right]-1) \, dx \right]  \times \frac{4}{3} \, \pi \!\! \int_{0}^{\infty} W(x) \, exp\left[-W(x) \, t\right] \, dx,
\label{non_dim_I_t}
\end{equation}

\subsection*{Intrinsic Point Defect Formation Enthalpies}

Doping of many wide-band gap semiconductors has proven to be challenging.\cite{walukiewicz2001} Figure~\ref{Defects_Flux}~(a) shows a heat map of $E_F$, which is determined by the charge balance between the predominant donor and acceptor point defects. The Fermi level lies above mid-gap over the majority of the single phase region where ZnSiP$_2$ is stable, resulting in  intrinsic to moderately \emph{n}-type conductivity. This result is promising as many compound semiconductors suffer from degenerate conductivity. In fact, ZnSiP$_2$ has been synthesized as both \emph{n}- or \emph{p}-type material, likely due to impurities arising from the synthesis techniques. Crystals grown in a Zn flux have been reported with both \emph{n}-type,\cite{siegel1974, humphreys1975, kuhnel1975, humphreys1976, siegel1976, kuhnel1978} and \emph{p}-type conductivity.\cite{ziegler1973, humphreys1976, gorban1976, siegel1976} All reports of growth in Sn flux have been \emph{n}-type.\cite{springthorpe1968, clark1973, shay1973, siegel1974, humphreys1976, gorban1976, siegel1976, kuhnel1978}  Halogen assisted vapor transport growth with I has produced \emph{n}-type crystals,\cite{clark1973, shay1973} while use of Cl as the carrier gas, with ZnCl$_2$ and PbCl$_2$ as Cl sources, has produced \emph{p}-type crystals.\cite{ ziegler1973, ziegler1974, gorban1976, siegel1976} There is one report of \emph{n}-type conductivity resulting from Cl vapor transport using SiCl$_4$ as the Cl source.\cite{siegel1974} The crystals have been intentionally, extrinsically doped \emph{n}-type by adding Se, Te, In, or Ga, and p-type by adding Cu. \cite{springthorpe1968, springthorpe1970, shay1973, ziegler1973, ziegler1978} The electronic properties that have been reported in these and other studies are given in Table \ref{electronic_optical}. When other elements are involved in the synthesis (Sn flux or I, or Cl vapor transport) they have been found as impurities in the resultant crystals.\cite{clark1973} These studies demonstrate that ZnSiP$_2$ can be synthesized either \emph{n}- or \emph{p}-type, using extrinsic dopants, with carrier concentrations acceptable for PV devices.

\vspace{24pt}

\begin{figurehere}
\centering
  \includegraphics[width=17.1cm]{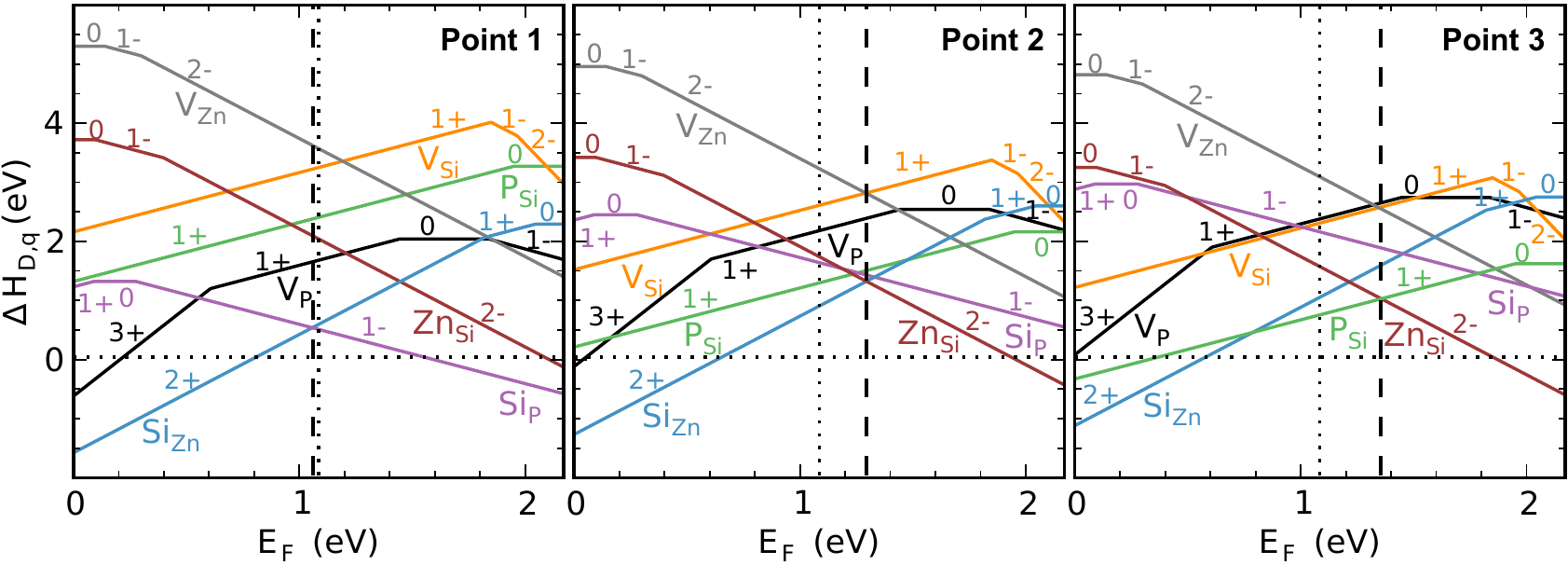}
  \caption{ Defect formation enthalpies ($\Delta$H$_{D,q}$) of 7 different point defects (vacancies and antisites) in all possible charge states (q ranging from 4- to 4+) at points 1, 2, and 3 on the phase diagram (a). The Fermi level is shown as a vertical dashed line along with mid-gap which is shown as a vertical dotted line.}
  \label{Defects_Flux_ESI}
\end{figurehere}

\subsection*{Stability of Photoelectrodes}
The stability of n-type semiconductors under aqueous anodic conditions has been a challenge for many photoelectrochemical applications (e.g. solar water splitting). The aqueous Fe(CN)$_6^{(3^-/4^-)}$ system has been used to monitor the water stability of many semiconductor anode materials with and without different protection layers. \cite{hu2014, shaner2015, strandwitz2013} When ZnSiP$_2$ electrodes were tested under aqueous conditions for long periods of time (2-8 hours), some dissolution of the semiconductor was observed by scanning electron microscopy (SEM) and the photocurrent was observed to decrease slightly while the open circuit potential ($E_\text{oc}$) remained relatively constant.   To confirm that the dissolution current of the ZnSiP$_2$ was negligible, an electrode was also tested in a non-aqueous electrochemical cell using the ferrocene/ferrocinium (Fc$^{+/0}$) redox couple (10~mM ferrocene, 0.05~mM ferrocinium, LiClO$_4$ supporting electrolyte in dry acetonitrile).\cite{Ritenour2012}
The photocurrent was very similar (within 10\%) between the aqueous (50/350 mM Fe(CN)$_6^{3-/4-}$) and non-aqueous measurements; this is consistent with the expected photon flux through the solution being nearly the same (within 1\%) in each case.  This similarity in photocurrent confirms that dissolution was not a significant component in the $J$-$E$ photoresponse observed for ZnSiP$_2$ electrodes under aqueous conditions.

\newpage

\twocolumn

\footnotesize{
\bibliography{ZnSiP2_Refs}
\bibliographystyle{rsc} 
}

\end{document}